\documentclass[a4paper]{aa}
\usepackage{graphicx,latexsym,amssymb,times}
\usepackage{psfig}
\def\gsim{ \lower .75ex \hbox{$\sim$} \llap{\raise .27ex \hbox{$>$}} } 
\def\lsim{ \lower .75ex\hbox{$\sim$} \llap{\raise .27ex \hbox{$<$}} } 

\begin{document}

\newcommand {\sax} {{\it Beppo}SAX }
\newcommand {\rosat} {{ROSAT }}
\newcommand {\rchisq} {$\chi_{\nu} ^{2}$} 
\newcommand {\chisq} {$\chi^{2}$}
\newcommand {\als} {$\alpha_1$}
\newcommand {\alh} {$\alpha_2$}
\newcommand {\ergs}[1]{$\times10^{#1}$ ergs cm$^{-2}$ s$^{-1}$}
\newcommand {\e}[1]{$\;\, \times10^{#1}$}

\title{Extreme Synchrotron BL Lac Objects }

\subtitle{ Stretching the Blazar sequence}

\author{
L. Costamante \inst{1,2}, G. Ghisellini \inst{1},
P. Giommi \inst{3}, G. Tagliaferri \inst{1}, A. Celotti \inst{4}, 
M. Chiaberge \inst{4}, G. Fossati \inst{5}, L. Maraschi \inst{6},
F. Tavecchio \inst{6}, A. Treves \inst{7}, A. Wolter \inst{6}
}

\offprints{L. Costamante; costa@merate.mi.astro.it}
\institute{{Osservatorio Astron. di Brera, via Bianchi 46 Merate, Italy;}
\and{Universit\`a degli Studi di Milano, via Celoria 16 Milano, Italy;}
\and{A.S.I., Science Data Center, c/o ESA--ESRIN, via G. Galilei, Frascati, Italy;} 
\and{SISSA/ISAS, via Beirut 2-4 Trieste, Italy;}
\and{CASS, University of California, San Diego, CA 92093-0494, USA;}
\and{Osservatorio Astron. di Brera, via Brera 28 Milano, Italy;}
\and{Universit\`a dell'Insubria, via Valleggio 11 Como, Italy.}}

 \date{Received December 01, 2000; accepted March 20, 2001}
 
 \authorrunning{L. Costamante et al.}

 \abstract{  
We performed an observational program with the X--ray satellite 
\sax, to study objects with extreme synchrotron peak frequencies 
($\nu_{\rm peak}>1$ keV). 
Of the seven sources observed, five revealed or confirmed their 
extreme nature. 
Four showed peak frequencies in the range 1--5 keV,
while one (1ES 1426+428) displayed a flat power law spectrum  
($\alpha_{x}=0.92\pm0.04$)
which locates its synchrotron peak at or above 100 keV.
This is the third source of this type ever found, after
Mkn 501 and 1ES 2344+514. 
In the context of the whole blazar class, the broad band properties of 
these objects confirm the scenario of a 
synchrotron peak smoothly spanning the IR -- X--ray range, which explains 
the multi--frequency properties of the blazar class.
Our data also  confirm the large $\nu_{\rm peak}$ variability 
which seems to characterize this class of sources, compared with lower 
$\nu_{\rm peak}$ objects.
Given the high synchrotron peak energies, which flag the presence of high 
energy electrons, these extreme BL Lacs are also good candidates for TeV 
emission, and therefore good probes of the IR background.
  }

\maketitle
%

\section{Introduction}

Among active galactic nuclei (AGNs), blazars are the objects 
most dominated by non thermal continuum emission, which extends from radio
to gamma rays, and whose properties are best explained by
emitting plasma  in relativistic motion towards the observer,
closely aligned with the line of sight (see e.g. Urry \& Padovani 1995). 
Their spectral energy distribution (SED), in a $\nu F_{\nu}$
representation, is characterized 
by two main broad peaks: that at low energy is commonly explained as 
synchrotron emission by highly relativistic electrons, while that
at higher energy is probably due to Inverse Compton emission of the same
electrons off seed photons, which can be either the same synchrotron photons
produced locally, or photons  from the external environment 
(see e.g. Ghisellini et al. 1996; Sikora et al. 1997; Ghisellini \&
Madau 1997; Dermer et al. 1997).

Among blazars, BL Lacertae objects are the sources  with the 
highest variety of synchrotron peak frequencies, ranging from the IR--optical 
to the UV--soft-X bands (called Low or High energy peak BL Lacs,
i.e. LBL or HBL, respectively; see Padovani \& Giommi 1995).
In the past the picture given by X--ray and radio surveys 
presented a bimodality in the distribution of the sources, 
with radio--selected objects mostly LBLs, and X--ray selected mostly HBLs.
New results are now revealing a more continuous scenario.
On one hand,
new surveys like the DXRBS (Perlman et al. 1998), the RGB
(Laurent--Muehleisen et al. 1998) and the REX (Caccianiga et al. 1999)
are showing a more continuous and ``smoother" range in peak frequencies. 
The ``gap" in the SEDs properties is now filled
thanks to the covering of previously unexplored regions of the parameters 
space (in $F_x$, $F_r$ and broad band
spectral indices $\alpha_{rx}$, $\alpha_{ro}$ and $\alpha_{ox}$). 
On the other hand, the \sax  observations of Mkn 501 (Pian et al. 1997)
and 1ES 2344+514 (Giommi et al. 2000) have revealed that  the 
synchrotron peak not only can reach the 2--10 keV  band 
(as was known for few sources, like Mkn 421), but  can go even further, 
up or above 100 keV, at least in a flaring state.
These exceptional observations have extended by nearly two orders of magnitude
the known ``available"  range of  synchrotron peak frequencies,
disclosing a new region of possible physical parameters  that can give 
us valuable insights on the still unknown particle acceleration mechanism.

At present, however, very little is known about such extreme objects:
we do not know if they are very rare or not,
and if these energies are reached only during flares or also  
in normal, ``quiescent" states.
Another interesting topic is if there exists an upper  limit to the
peak frequency, and if we have already observed it or not. 
In order to answer to these questions, a necessary step is to find and
study other objects of this kind, sampling more accurately the high 
energy region ($>1$ keV) of the peak sequence.
With this goal,we have proposed and performed an observational campaign 
using the \sax  satellite (Boella et al. 1997), 
after selecting a sample of candidates. 
The X--ray band  is ideal for this purpose: 
in this band, objects characterized by different
synchrotron peak frequencies show  different spectral properties,
according to the dominant radiation process (synchrotron or inverse Compton).
This provides means to roughly estimate the position
of the synchrotron peak, with the aid of the overall SED shape
to recognize the origin of the X--ray emission.

In fact, the X--ray spectral index might even be considered {\it the} 
parameter by which to classify the source:
BL Lacs which have the synchrotron peak in the IR band (i.e. LBLs), have
a flat ($\alpha_x<1$) X--ray spectral index, due to inverse 
Compton emission, as for FSRQs
\footnote{
We call flat or steep a power law spectrum with index
$\alpha<1$ or $\alpha>1$ respectively.
This convention corresponds to rising or decaying (with frequency)
spectra in a $\nu$--$\nu F_\nu$ plot.
}.
As the peak shifts towards higher energies, the synchrotron emission 
becomes more and more dominant in the X--ray band, and so we have, 
in sequence: intermediate BL Lacs, where we see the high energy tail 
of the synchrotron emission in the soft X--ray band and the
flat inverse Compton spectrum in the hard band, resulting in a concave 
X--ray spectral shape ($\alpha_{\rm soft}>1$ and $\alpha_{\rm hard}<1$); 
HBLs, where the X--ray band is completely dominated by the synchrotron 
emission above its peak, and so show a steep spectral index ($\alpha_x>1$); 
and Extreme BL Lacs, where the synchrotron peak becomes visible directly 
{\it in} the X--ray band at energies $>1$ keV (resulting in a convex spectrum 
with $\alpha_{\rm soft}<1$  and $\alpha_{\rm hard}>1$, or in a flat
X--ray slope all the way up to the maximum observable X--ray energies,
as in the case of Mkn 501).
Thanks to its wide energy band (0.1--100 keV), the  \sax satellite offers   
the best opportunity to identify and study these extreme sources.

\section{The sample}

Our candidates have been selected from the HEAO1 survey,
the Einstein Slew Survey and the Rosat All Sky Survey
Bright Sources Catalogue (RASSBSC), using also information in other bands
to construct the  SEDs and to determine the broad band spectral indices
($\alpha_{ro}$, $\alpha_{ox}$ and $\alpha_{rx}$).
These are good indicators of the location of the synchrotron peak,
since objects of different characteristics tend to gather in different regions 
of the parameter space (Stocke et al. 1991; Padovani \& Giommi 1995; 
Fossati et al. 1998; Padovani et al. 1997; see also Fig. 1).  

The selection criteria were based on properties suggesting a high 
value of the synchrotron peak energy:
\begin{itemize}
\item a very high $F_x/F_{\rm radio}$ ratio ($>3\times10^{-10}$ erg cm$^{-2}$
s$^{-1}$/Jy, measured at least once, in the [0.1--2.4] keV band and at 
5 GHz respectively);
\item flat X--ray spectrum (when available), connecting smoothly with
the flux at lower frequencies;
\item appropriate values of $\alpha_{ro}$, $\alpha_{ox}$ and $\alpha_{rx}$
(see e.g. Fig. 1).
\end{itemize}
In addition, a high X--ray flux ($>10^{-11}$ erg s$^{-1}$ cm$^{-2}$)
was also requested, to achieve a good detection in the PDS instrument.
\begin{figure}
\hspace{-0.6cm}
\psfig{figure=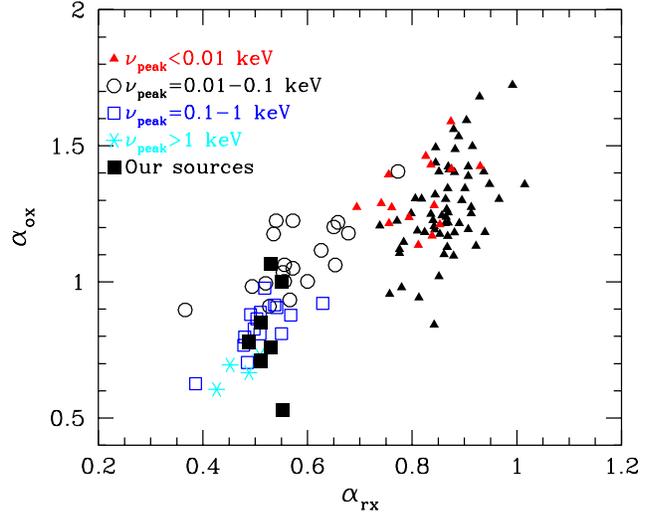,angle=0,width=9.5cm}
\vskip -2 true cm
\caption{The plane $\alpha_{ox}$--$\alpha_{rx}$ for the SLEW sample of BL Lacs,
the 1 Jy sample of BL Lacs and the 2 Jy sample of FSRQs. 
All sources with $\alpha_{rx}<0.7$ are X--ray selected BL Lacs. 
The SED of all sources has been fitted with a cubic function in order to find 
the synchrotron peak frequency $\nu_{\rm peak}$ (Fossati et al. 1998). 
Filled squares mark the location of the observed candidates.}  
\end{figure}

\section{{\sl{\bf Beppo}}SAX observations}

In Table 1 we report the observation log and some basic information for 
the seven observed sources. 
The LECS and PDS exposure times are usually shorter than the MECS ones,
due to light leakage in the LECS instrument, which forces to take the exposures
only during  the part of the orbit in the earth shadow cone; 
and due to the on--off source switching for the PDS instrument. 

The data analysis was based on the linearized, cleaned event files obtained
from the \sax~ Science Data Center (SDC) on--line archive 
(Giommi \& Fiore 1997).
The data from the single MECS units were merged in one event file by SDC, 
based on sky coordinates. 
The event file was then screened with a time filter (from SDC) to
exclude those intervals related to events without attitude solution
(i.e. conversion from detector to sky coordinates), as recommended by SDC
(see Handbook for NFI Spectral Analysis, Fiore et al., 1999).
Due to remaining calibration uncertainties, the channels 1--10 and over 4 keV
for the LECS, and 0--36 and 220--256 for the MECS were excluded from the 
spectral analysis. 
Standard extraction radii of 4 and 8 arcmin  for MECS and LECS respectively were used, 
except for PKS 0548--322 (see paragraph 3.2.3). 

The PDS instrument has no imaging capability, and its f.o.v.
($\sim1.4^{\circ}$ FWHM) is larger than the other Narrow Field Instruments 
(radius $\sim28^{\prime}$ for the MECS). 
Therefore there is the possibility for PDS spectra to be
contaminated by hard serendipitous sources in the f.o.v, not visible
in the smaller MECS images. 
To explore this possibility, for each object we checked in the 
NED and WGACAT databases for the presence of potentially contaminating sources.

\begin{table}
\centering
\caption{\bf Log of the observations}
\begin{tabular}{lllc}
\hline
\hline  \vspace*{-2.5mm}\\
Date   &  Detector  &  Exp. Time  & Net countrate full band   \\
        &          &      seconds    &    counts/s   \\
\hline
\hline  \vspace*{-1.5mm}\\
\multicolumn{4}{c}{\bf 1ES 0033+595 \hspace*{1cm} {\bf z= ?
\hspace*{1cm} gal. N$_{\bf\rm H}$= 4.24 x10$^{\bf21}$ cm$^{-2}$ } } \\
\hline
\vspace*{-3mm} \\
18--19  Dec. 99    & LECS   &  13267   &  0.368 \,$\pm0.006$ \\
                    & MECS   &  43424  &  0.766 \,$\pm0.004$ \\
                    & PDS    &  22877  &  0.503 \,$\pm0.052$ \\                    
\hline
\hline  \vspace*{-1.5mm}\\
\multicolumn{4}{c}{\bf 1ES 0120+340 \hspace*{1cm} {\bf z= 0.272
 \hspace*{0.5cm} gal. N$_{\bf\rm H}$= 5.15 x10$^{\bf20}$ cm$^{-2}$} } \\
\hline
\vspace*{-3mm} \\
3--4 Jan. 99      & LECS   &  9661   &  0.198 \,$\pm0.005$ \\
                    & MECS   &  32039  &  0.260 \,$\pm0.003$ \\
                    & PDS    &  16263  &  0.381 \,$\pm0.065$ \\
                    & & & \\
2--3 Feb. 99      & LECS   &  4848   &  0.179 \,$\pm0.008$ \\
                    & MECS   &  21562  &  0.213 \,$\pm0.003$ \\
                    & PDS    &  10128  &  0.196 \,$\pm0.082$ \\
\hline
\hline  \vspace*{-1.5mm}\\
\multicolumn{4}{c}{\bf PKS 0548--322 \hspace*{1cm} {\bf z= 0.069
 \hspace*{0.5cm} gal. N$_{\bf\rm H}$= 2.51 x10$^{\bf20}$ cm$^{-2}$ } } \\
\hline
\vspace*{-3mm} \\
20 Feb. 99        & LECS   &  9328   &  0.280 \,$\pm0.006$ \\
                    &  MECS  &  12439  &  0.350 \,$\pm0.005$ \\
                    &  PDS   &  9355   &  0.142 \,$\pm0.083$ \\
                    & & & \\
26--27 Feb. 99    & MECS   &  2025   &  0.309 \,$\pm0.013$ \\
                    & & & \\
7--8 April 99     & LECS   &  5251   &  0.222 \,$\pm0.008$ \\
                    & MECS   &  18943  &  0.248 \,$\pm0.004$ \\
                    & PDS    &  10164  &  0.039 \,$\pm0.085$ \\
\hline
\hline  \vspace*{-1.5mm}\\
\multicolumn{4}{c}{\bf GB 1114+203 \hspace*{1.1cm} {\bf z= ?
\hspace*{1cm} gal. N$_{\bf\rm H}$= 1.36 x10$^{\bf20}$ cm$^{-2}$ }  } \\
\hline
\vspace*{-3mm} \\
 13--15 Dec. 99     & LECS   &  13551  &  0.195 \,$\pm0.004$ \\
                    & MECS   &  42355  &  0.128 \,$\pm0.002$ \\
                    & PDS    &  20432  &  -0.026 \,$\pm0.051$ \\
\hline
\hline  \vspace*{-1.5mm}\\
\multicolumn{4}{c}{\bf 1ES 1218+304 \hspace*{1cm} {\bf z= 0.182
\hspace*{0.5cm} gal. N$_{\bf\rm H}$= 1.78 x10$^{\bf20}$ cm$^{-2}$ }  } \\
\hline
\vspace*{-3mm} \\
 12--14 July 99     & LECS   &  10626  &  0.284 \,$\pm0.006$ \\
                    & MECS   &  42693  &  0.268 \,$\pm0.003$ \\
                    & PDS    &  20670  &  0.108 \,$\pm0.051$ \\                      
\hline
\hline  \vspace*{-1.5mm}\\
\multicolumn{4}{c}{\bf 1ES 1426+428 \hspace*{1cm} {\bf z= 0.129
 \hspace*{0.5cm} gal. N$_{\bf\rm H}$= 1.36 x10$^{\bf20}$ cm$^{-2}$ }  } \\
\hline
\vspace*{-3mm} \\
8--9 Feb. 99      & LECS   &  15153  &  0.232 \,$\pm0.004$ \\
                    & MECS   &  40657  &  0.293 \,$\pm0.003$ \\
                    & PDS    &  20432  &  0.487 \,$\pm0.058$ \\
\hline
\hline \vspace*{-1.5mm}\\
\multicolumn{4}{c}{\bf H 2356--309  \hspace*{1.2cm} { \bf z= 0.165
 \hspace*{0.5cm} gal. N$_{\bf\rm H}$= 1.33 x10$^{\bf20}$ cm$^{-2}$ }} \\
\hline
\vspace*{-3mm} \\
21--22 June 98    &  LECS  &  15098  &  0.327 \,$\pm0.005$ \\
                    &  MECS  &  41034  &  0.374 \,$\pm0.003$ \\
                    &  PDS   &  18035  &  0.150 \,$\pm0.063$ \\
\hline
\hline  
\end{tabular}
\end{table}

\subsection{Time analysis}

We looked for time variability in each observation, using time bins
from 500 to 10,000 s: no significant variability has been detected 
(\chisq--test constance probabilities $>20 \%$), except for 1ES 1426+428.
For this source the lightcurve clearly shows a trend (see Fig. 2),
with a flux decreasing  in the central part of the observation
by about $\sim10\%$ in 10 hours, before rising again.  
This variability is indicated by the low \chisq--test  
probability of constant flux ($<0.02\%$), and confirmed by the fact that
the same pattern is visible in the lightcurves from the single
MECS2 and MECS3 instruments.
No significant variations are present in the LECS lightcurves,
but the lower statistics does not allow to draw conclusions on a different
behaviour of the low energy emission.
\begin{figure}
\psfig{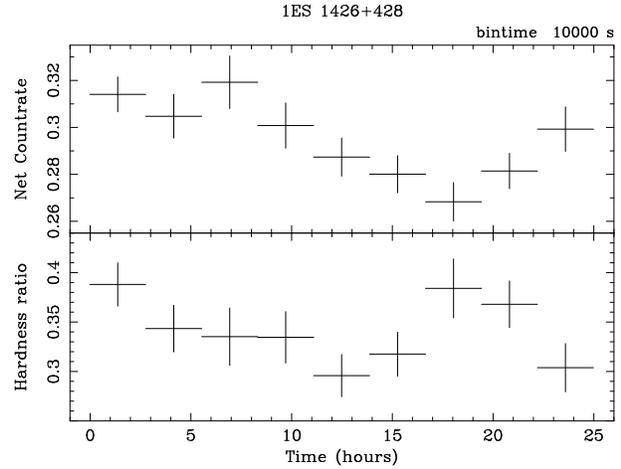}
\vskip 0.5 true cm
\caption{MECS lightcurve and Hardness Ratio for 1ES 1426+428, defined as 
(5--10 keV)/(2--5 keV) countrates. }
\end{figure}
Fig. 2 also shows the corresponding hardness ratio (HR): 
the scatter is not compatible with a constant value.
After the 11th hour there seems to be an anticorrelation between the 
HR and the countrate.
However the significance of this anticorrelation is small: there is still a 
$\sim$20\% probability to obtain the same correlation coefficient from 
a random sample.

\begin{table*}[!ht]
\vspace{1.2cm}
\centering
{ Single power-law fits, $\;$ Lecs+Mecs}
\begin{tabular}{lllllllll}
\vspace*{-1mm} \\
\hline
\hline
\vspace*{-2mm} \\
Source & Date  & $N_{\rm H}$ & $\alpha_x$ & $F_{1~{\rm keV}}$ &
$F_{2-10~{\rm keV}}$ & Lecs & $\chi^2_r$/d.o.f. & notes \\
\cline{7-7}
{\footnotesize (IAU name)} &  & $10^{20}$ cm$^{-2}$ &  & $\mu$Jy &
ergs cm$^{-2}$s$^{-1}$ &   Mecs &  & 
 \vspace*{1mm} \\
\hline
\hline
\vspace*{-3mm} \\
{\bf 0033+595} &  18/12/99 & $ 42.4$\, {\footnotesize{\it fixed}} &  $0.93\pm0.02$ &
$13.5\pm0.3$ &$5.9$ \e{-11} & 0.60 & 2.24/152 &  $+$ PDS \\
      &  &  $75\pm7$ & $1.05\pm0.04$ &
$16.8\pm0.7$ & $6.0$ \e{-11} & 0.69 & 1.12/151 &  
\vspace*{1mm}\\
\hline
\vspace*{-3mm} \\
{\bf 0120+340} & {\bf 3/1/99} & $ 5.15$\, {\footnotesize{\it fixed}} &  $1.08\pm0.04$ &
$5.0\pm0.3$ & $1.7$ \e{-11} & 0.68 & 0.98/122 & LECS$>0.3$ keV$^{1)}$ \\
      &  &  $8.5^{-3.9}_{+4.3}$ & $1.11\pm0.06$ &
      $5.2\pm0.3$ & $1.7$ \e{-11} & 0.70 & 0.96/121 &  LECS$>0.3$ keV$^{1)}$ \\
      &  &  $9.9^{-1.9}_{+3.0}$ & $1.13\pm0.06$ &
$5.2\pm0.3$ & $1.7$ \e{-11} & 0.71 & 0.96/122 &  LECS total$^{1)}$
\vspace*{2mm}\\
        &  2/2/99  & $ 5.15$\, {\footnotesize{\it fixed}} &  $1.23\pm0.05$ & $4.8\pm0.3$ &
$1.3$ \e{-11} & 0.63 & 1.15/95 &  \\
    &     &    $8.9^{-2.4}_{+6.1}$ & $1.31\pm0.09$ &
$5.4^{-0.4}_{+0.5}$ & $1.3$ \e{-11} & 0.68 & 0.96/94 & 
\vspace*{1mm}\\
\hline
\vspace*{-3mm} \\
{\bf 0548--322} &  20/2/99  & $ 2.51$\, {\footnotesize{\it fixed}} &  $0.88\pm0.04$ & $5.1\pm0.3$ &
$2.4$ \e{-11} & 0.77 & 1.83/85 &  \\
      &    & $4.9^{-0.8}_{+1.1}$ & $1.03\pm0.07$ &
$6.3\pm0.4$ & $2.3$ \e{-11} & 0.73 & 1.09/84 & 
\vspace*{2mm}\\
        &  {\bf 26/2/99}  & $ 2.51$\, {\footnotesize{\it fixed}} &  $1.32\pm0.19$ &
$7.5^{-1.5}_{+1.9}$ & $1.8$ \e{-11} & --- & 0.58/17 & MECS only
\vspace*{2mm}\\
      & {\bf 7/4/99}   &  $ 2.51$\, {\footnotesize{\it fixed}} &  $1.14\pm0.05$ & $5.0\pm0.3$ &
$1.6$ \e{-11} & 0.66 & 1.10/79 &  \\
      &          & $4.7^{-1.0}_{+1.6}$ & $1.24\pm0.08$ &
$5.7\pm0.4$ & $1.5$ \e{-11} & 0.68 & 0.72/78 & 
\vspace*{1mm}\\
\hline
\vspace*{-3mm} \\
{\bf 1114+203}  &  13/12/99  &   $1.36$\, {\footnotesize{\it fixed}} & $1.53\pm0.03$ & 
$3.7\pm0.2$ & $6.7$ \e{-12} & 0.74 & 3.14/83 &  \\
       &           & $3.1\pm0.4$ & $1.82^{-0.07}_{+0.05}$ &
$5.2^{-0.3}_{+0.2}$ & $6.4$ \e{-12} & 0.65 & 1.42/82 &
\vspace*{1mm}\\
\hline
\vspace*{-3mm} \\
{\bf 1218+304}  &  12/7/99  &  $1.78$\, {\footnotesize{\it fixed}} & $1.39\pm0.03$ &
$7.1^{-0.3}_{+0.2}$ & $1.6$ \e{-11} & 0.66 & 1.95/94 &  \\
       &           & $3.0^{-0.3}_{+0.4}$ & $1.51\pm0.04$ &
$8.1\pm0.3$ & $1.5$ \e{-11} & 0.67 & 1.15/93 &              
\vspace*{1mm}\\
\hline
\vspace*{-3mm} \\
{\bf 1426+428}  & {\bf 8/2/99}  &   $1.36$\, {\footnotesize{\it fixed}} & $0.91\pm0.03$ & $4.6\pm0.2$ &
$2.0$ \e{-11} & 0.68 & 1.00/90 &  \\
       &           & $1.5^{-0.3}_{+0.4}$ & $0.92\pm0.04$ &
$4.6\pm0.2$ & $2.0$ \e{-11} & 0.68 & 1.00/89 &
\vspace*{1mm}\\
\hline
\vspace*{-3mm} \\
{\bf 2356--309}  &  21/6/99  &    $1.33$\, {\footnotesize{\it fixed}} & $0.96\pm0.02$ & 
$6.1\pm0.2$ &
$2.5$ \e{-11} & 0.72 & 3.17/37 &  $+$ PDS  \\
          &          & $2.3\pm0.3$ & $1.06\pm0.04$ &
$7.0\pm0.3$ & $2.5$ \e{-11} & 0.70 & 1.58/36 &
\vspace*{1mm}\\
\hline
\hline
\multicolumn{9}{l}{\footnotesize{Errors at 90\% conf. level for 1 
(fixed $N_{\rm H}$) and 2 (free $N_{\rm H}$) par. of interest. $\;\; ^{1)}$see text   }} \\
\end{tabular}
\caption{Single power--law fits with galactic and free absorption. 
In bold are highlighted the dates of those observations for which the single power-law model
provides an adequate representation of the data. For the other observations, instead,
a significantly better fit is provided by the broken power-law models 
(see Table 3).}
\end{table*}

\subsection{Spectral Analysis}
The spectral analysis was performed with the XSPEC 10.0 package,
using the latest available response matrices and blank--sky background
files (at 01/2000). 
\begin{table*}[!ht]
\centering
{Broken power--law  fits }
\begin{tabular}{lllllllll}
\vspace*{-1mm}\\
\hline
\hline
\vspace*{-3mm} \\
   $N_{\rm H}$ & $\alpha_1$ & $E_{\rm break}$ & $\alpha_2$ &$F_{1~ \rm keV}$ 
    &$F_{2-10~\rm keV}$ & Lecs & $\chi^2_r$/d.o.f. & F--test  \\
\cline{7-7}
\vspace*{1mm}
 $10^{20}$ cm$^{-2}$ & & keV & & $\mu$Jy &  ergs cm$^{-2}$s$^{-1}$ &
   Mecs & &
\vspace*{1mm}\\
\hline
\hline
\vspace*{-3mm} \\
\multicolumn{9}{l}{{\bf 0033+595} $\;\;$ {\footnotesize fits LECS+MECS+PDS } }      \\
\hline
\vspace*{-3mm} \\
$42.4$ \,{\it \footnotesize fixed} & $0.34\pm0.13$ & $2.4\pm0.2$ &
$ 1.03\pm0.03$ & $8.7\pm0.7$ &  $5.9$ \e{-11} & 0.70 & 1.12/150 & $\gsim68\%$
\vspace*{1mm}\\
$62^{-11}_{+13}$ & $0.8^{-0.3}_{+0.2}$ & $2.9^{-0.6}_{+2.5}$ &
 $1.08^{-0.05}_{+0.12}$ & $13.2\pm2.5$ & $6.0$ \e{-11} & 0.70 & 1.05/149 & $\gsim 99.6\%$
\vspace*{1mm} \\
\hline
\vspace*{-3mm} \\
\multicolumn{9}{l}{{\bf  0120+340} $\;\;$observation of 2/2/99} \\
\hline
\vspace*{-3mm} \\
 $5.15$ \,{\it \footnotesize fixed} & $0.8^{-0.9}_{+0.3}$ & $1.4^{-0.6}_{+0.8}$ &
$ 1.32\pm0.08$ & $4.5^{-0.4}_{+1.9}$ &  $1.3$ \e{-11} & 0.70 & 0.92/93 &
$\gsim 97\%$
\vspace*{1mm}\\
 $5.6^{-4.4}_{+4.9}$ & $0.8^{-1.7}_{+0.3}$ & $1.4^{-0.4}_{+6.9}$ &
 $ 1.32\pm0.09$ & $4.5^{-0.5}_{+1.3}$ & $1.3$ \e{-11} & 0.70 & 0.93/92 & $\gsim 92\%$
\vspace*{1mm}\\
\hline
\vspace*{-3mm} \\
\multicolumn{9}{l}{{\bf  0548--322} $\;\;$ observation of 20/2/99} \\
\hline
\vspace*{-3mm} \\
 $2.51$ \,{\footnotesize \it fixed} & $0.3^{-0.3}_{+0.2}$ & $1.0^{-0.2}_{+0.4}$ &
 $ 1.04\pm0.05$ & $6.5^{-0.8}_{+0.9}$ &  $2.3$ \e{-11} & 0.73 & 1.04/83 & $\gsim 97\%$
\vspace*{1mm}\\
$4.2^{-0.9}_{+1.1}$ & $0.91^{-0.14}_{+0.08}$ & $4.5^{-2.3}_{+1.7}$ &
$1.4^{-0.3}_{+0.6}$ & $5.7\pm0.3$ & $2.3$ \e{-11} & 0.77 & 0.95/82 & $\gsim 99.8\%$
\vspace*{1mm}\\
\hline
\vspace*{-3mm} \\
\multicolumn{9}{l}{{\bf 1114+203} } \\
\hline
\vspace*{-3mm} \\
$1.36 \,$ {\footnotesize \it fixed}  & $1.37^{-0.07}_{+0.06}$ & $1.6^{-0.3}_{+0.4}$ &
 $1.89\pm0.08$ & $4.4\pm0.4$ &  $6.2$ \e{-12} & 0.77 & 0.93/62 & $>99.9 \%$
\vspace*{1mm}\\
$1.5\pm0.6$ & $1.40^{-0.22}_{+0.18}$ & $1.6^{-0.4}_{+0.7}$ &
$1.89\pm0.09$ & $4.4\pm0.4$ & $6.2$ \e{-12} & 0.77 & 0.95/61 & $>99.9 \%$
\vspace*{1mm}\\
\hline
\vspace*{-3mm} \\
\multicolumn{9}{l}{{\bf 1218+304} } \\
\hline
\vspace*{-3mm} \\
$1.78 \,$ {\footnotesize \it fixed}  & $1.15^{-0.08}_{+0.07}$ & $1.7^{-0.3}_{+0.4}$ &
$1.56\pm0.05$ & $7.0^{-0.5}_{+0.6}$ &  $1.5$ \e{-11} & 0.72 & 0.84/92 & $>99.9 \%$
\vspace*{1mm}\\
$1.6\pm0.7$ & $1.11^{-0.23}_{+0.19}$ & $1.6^{-0.4}_{+0.5}$ &
$1.56\pm0.06$ & $6.9\pm0.6$ & $1.5$ \e{-11} & 0.77 & 0.85/91 & $>99.9 \%$
\vspace*{1mm}\\
\hline
\vspace*{-3mm} \\
\multicolumn{9}{l}{{\bf 2356--309} $\;\;$ fits {\footnotesize LECS+MECS+PDS}} \\
\hline
\vspace*{-3mm} \\
$1.33 \,$ {\footnotesize \it fixed}  & $0.78^{-0.07}_{+0.06}$ & $1.8^{-0.4}_{+0.4}$ &
 $ 1.10\pm0.04$ & $6.2\pm0.2$ &  $2.5$ \e{-11} & 0.73 & 0.95/35 & $>99.9 \%$
\vspace*{1mm}\\
$1.4\pm0.5$ & $0.80^{-0.18}_{+0.13}$ & $1.8^{-0.5}_{+0.8}$ &
$1.10\pm0.05$ & $6.2\pm0.2$ & $2.5$ \e{-11} & 0.73 & 0.97/34 & $>99.9 \%$
\vspace*{1mm}\\
\hline
\hline
\vspace*{-3mm} \\
\multicolumn{9}{l}{\footnotesize  Errors at 68\% conf. level for 3 and 4 parameters
of interest ($\Delta\chi^2=3.52$ and 4.72 respectively).  
 } \\
\end{tabular}
\caption{Broken power--law fits with galactic and free absorption. 
For these datasets, the broken power-law
model is statistically preferred with respect to a single power-law model   
(the F--tests in the last column are against
 the single power--law model with free absorption).}
\end{table*}    
Particular attention was paid to  
the LECS  blank--sky files, chosing those
extracted in the same coordinate frame as the source files,
to avoid the errors in the background subtraction arising from the different
scaling of the area between raw and sky--detector coordinates.

The spectra were rebinned  both with more than 20 
net counts in each new bin and using the rebinning files provided 
by SDC, applying the Gehrels statistical weight (Gehrels 1986) when
the resulting net counts were below 20 (typically 12--15 in some low energy
channels of LECS). 
Various checks have shown that our results are independent,
within the uncertainties, of the adopted rebinning.
The LECS/MECS normalization factor was let free to vary: normal acceptable
values should be within the 0.65--1.0 range, as indicated by SDC.
For the PDS instead, 
the normalization factor was fixed at 0.9, as results from 
intercalibration tests performed on known sources (Fiore et al. 1999).

In order to unveil the possible presence of the synchrotron peak 
in the observed band,
we have fitted routinely all  datasets both  with single and broken 
power--law models, with fixed Galactic and free absorption. 
The absorbing column density was parameterized in terms of $N_{\rm H}$, 
the HI column density, with heavier elements fixed at solar abundances 
and cross sections taken from Morrison \& McCammon (1983). 
The Galactic value was derived from the {\tt nh} program at HEASARC 
(based on Dickey \& Lockman 1990), except for 1ES 1218+304, for which a direct
measure is available (Elvis et al. 1989).
The results from  the single power-law fits are reported in 
Table 2, while  Table 3 reports the 
broken power-law fits parameters.
Adopting a conservative approach,
we assessed the presence of a spectral break when the broken power-law model 
was 
significantly better, according to the F--test, even with respect to the single 
power--law model with free absorption. 
These are the cases presented in Table 3.
Instead, when the fits were not significantly improved,
we considered the single power--law model with free or 
galactic absorption as an adequate representation of the data.

For most of the single power--law models, as shown in Table 2, 
the fits are improved (with significance $>99.9\%$)
adopting an absorption value greater than the galactic one.
In most cases, however,
the broken power--law models with galactic absorbing columns
provided even better fits (according to the F--test, see Table 3).
For 1ES 0033+595 and PKS 0548--322, instead, there is evidence 
for a higher column density also using the broken power--law models.
To check if this extra absorption is real, we have used for
these datasets also the ``curved spectrum" model, 
as in Fossati et al. 2000. 

In the following sections we present the details of the analysis
for each object, 
along with some notes on the individual characteristics.

\subsubsection{1ES 0033+595}
\begin{figure}
\centering
\resizebox{7.5cm}{!}{\includegraphics[angle=-90]{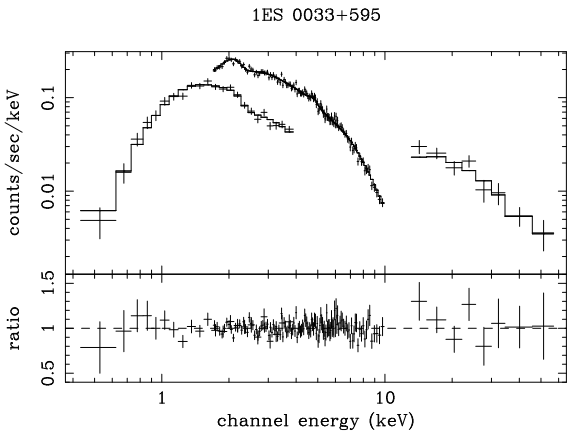}}
\vspace{0.2cm}
\caption{Broken power-law fit to the LECS+MECS+PDS data of 1ES 0033+595, 
with free galactic absorption.}
\end{figure}
This source was  detected in the LECS instrument only above 0.4 keV,
due to the high galactic absorption, and in the PDS instrument up to $\sim60$ keV.
The spectrum is characterized by a curved shape,
which is 
better fitted by a broken power-law than by a single power-law model
(with both fixed and free absorption). 
The best fit, however, still requires a column density in excess of the galactic value
(see Table 3), with high significance (F--test $\gsim 99.8\%$). 
The spectral indices before and after the break energy  
are lower and higher than 1, respectively, thus locating  the 
synchrotron peak at the break energy itself, around 3 keV.
The PDS data agree well with the extrapolation of the power-law 
after the break.  
The obtained $N_{\rm H}$ values higher than the galactic one cannot be 
accounted for by a curved model:
this gives unplausible values for the spectral index at low energies
($\alpha_1=0.1$ at 0.5 keV), for galactic absorption.
More likely values ($\alpha_1=0.6-0.8$) are obtained with a column density
close to the one given by the broken power-law fit ($5.7-6.3$ x10$^{21}$ cm$^{-2}$).    
This object has been observed also by the Hubble Space Telescope as part of the ``snapshot
survey" of BL Lac objects, and was resolved into 
two point-like sources,
with a separation of $1\arcsec.58$.
It is a possible case of gravitational lensing
(for a discussion of this hypothesis, see Scarpa et al. 1999, and
Falomo \& Kotilainen 1999).
\begin{figure*}
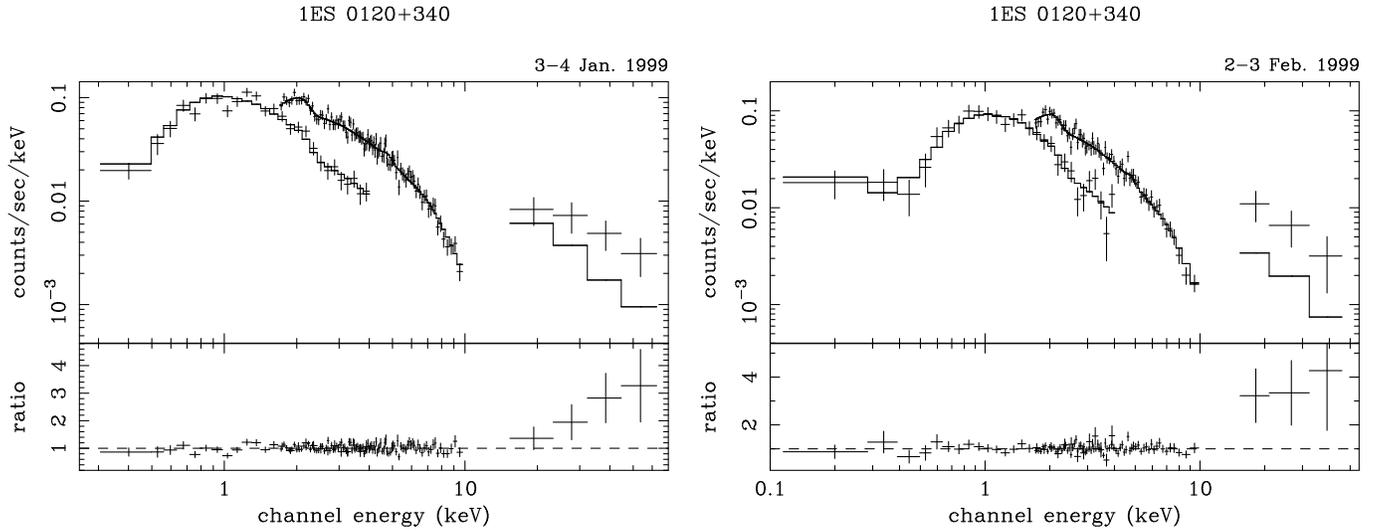

\resizebox{\hsize}{!}{\includegraphics[angle=-90]
{H2568F4L.PS}\hspace{1.0cm}\includegraphics[angle=-90]{H2568F4R.PS}}
\caption{Best fits data and folded model, plus residuals, for the two 
observations of 1ES 0120+340. On the left, the model is a single
power--law + galactic absorption (see text); on the right a broken power--law + galactic 
absorption.  PDS data are probably contaminated (see text), and are not used 
for the fits.}
\end{figure*}

\subsubsection{1ES 0120+340} 
\begin{figure}
\vspace{0.5cm}
\centering
\resizebox{7.5cm}{!}{\includegraphics[angle=-90]{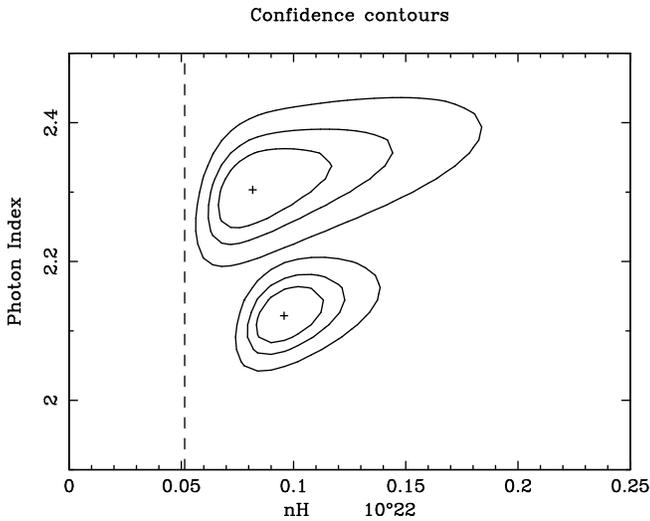}}
\vspace{0.5cm}
\caption{Confidence contours for the photon index and column density from
the single power-law fits to 1ES 0120+340 data. The upper one is for the
2/2/99 observation, the lower for the 3/1/99 one. The line indicates the galactic
column density.}
\end{figure} 
This source was observed twice in a month,
and was about $\sim20\%$ weaker during the second observation.
The best fits are shown in Fig. 4.
In both epochs, a good fit is obtained with the single power--law model,
but with a free absorption well in excess
of the galactic value, at more than
99.9\% confidence (see Fig. 5). 
As shown in Fig. 5 (and also indicated by the different ratio
of MECS/LECS countrates), the spectral index has varied significantly,
so it was not possible to sum the observations to enhance the S/N ratio.

In the second observation, a better fit is provided by 
a broken power--law model with galactic absorption, 
according to the F--test ($\gsim 97\%$).
This is also consistent with previous results by
Brinkmann et al. 1994 (RASS data), which reported 
absorption values near the galactic one
[$\alpha= 0.93\pm 0.35$, $N_{\rm H}= 5.1\pm 1.4 \times10^{20}$ cm$^{-2}$, 
\rchisq$= 0.97(13\: d.o.f.)$, $F_x= 4.50\times 10^{-11}$ 
erg s$^{-1}$ cm$^{-2}$ in the ROSAT band]. 
In the first observation, instead, a higher column density is required
even with a broken power--law model 
[\rchisq$=$0.98($120\:d.o.f.$), to be compared with 
\rchisq$=$1.10($121\:d.o.f.$) found with galactic N$_{\rm H}$].
Comparing the two LECS datasets, the main differences lie
in the 0.1--0.3 keV range, where oddly the higher flux observation 
presents a lack of soft photons with respect to the lower flux data.
A check on the local background, compared to the blank--sky one, 
showed no relevant differences. 
Although this behavior could in principle be due to a variable absorption, 
either local or at the source, the low S/N in that energy range  
does not allow to draw reliable conclusions.
Beyond 0.3 keV, the galactic column density provides good fits both with
single and broken power--law models.
Given the previous ROSAT results, we regard it as the most 
likely hypothesis, however future observations with higher S/N are 
necessary to settle this issue.

\begin{figure*}
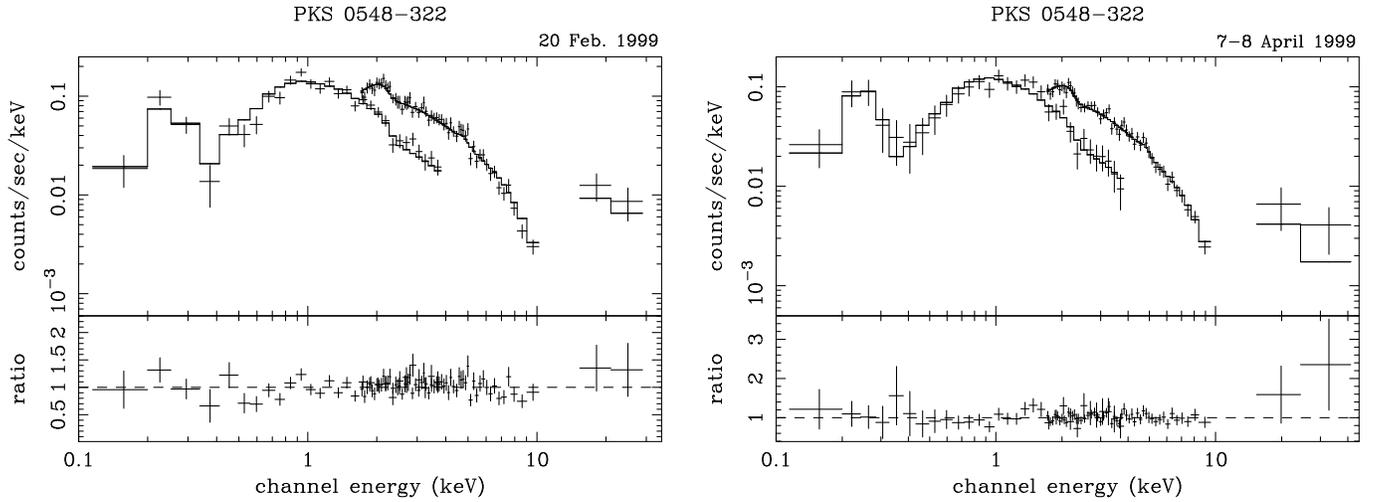

\resizebox{\hsize}{!}{\includegraphics[angle=-90]
{H2568F6L.PS}\hspace{1.5cm}\includegraphics[angle=-90]{H2568F6R.PS}}
\caption{Best fit data and folded model, plus residuals, for the first and
third observations of PKS 0548--322. 
The left panel shows the best fit model using a broken
power--law model, the right panel using a single power--law, 
both with free absorption.}
\end{figure*}

The source was detected also with the PDS (see Fig. 4), but with 
a flux incompatible with that expected by a simple extrapolation 
of the LECS+MECS data. 
Possible explanations are a flattening ($\alpha\sim1$) of the spectrum
due to the emerging of an inverse Compton component
(not unconceivable, given its SED, see Fig. 14), 
or the contamination by other sources in the field of view 
not visible in the imaging instruments. 
This last hypothesis is corroborated by the presence, at $36^{\prime}$ 
from the BL Lac object, of the galaxy NGC 513 (z=0.0195), identified as a 
Seyfert 2. 
Objects of this class, being heavily absorbed, can actually have a 
significant hard X--ray emission, even if the soft 
X--ray flux is low (the ROSAT WGACAT full band countrate for NGC 513 
is $0.0097\pm0.0014$ cts/s, 
corresponding to 1.45 $\times 10^{-13}$ erg cm$^{-2}$ s$^{-1}$ for 
galactic absorption).
To check this possibility, we have also added a power--law component to 
the model for the PDS data of 1ES 0120+340, to mimic the Seyfert 2 
emission in the hard band, after accounting for the off--axis efficiency 
of the instrument.
In this way, the PDS data of both epochs can be easily explained by a 
power--law component of $\alpha\sim 0.7$ with a normalization
at 1 keV of $2.5\pm1.4$  $\mu$Jy, which requires a column  density
of $\approx 4\times10^{22}$ cm$^{-2}$ to account for the ROSAT countrate.
These values are quite common for Seyfert 2 objects, 
so this is a likely possibility.

\subsubsection{PKS 0548--322}
\begin{figure}
\vspace{0.5cm}
\centering
\resizebox{7.5cm}{!}{\includegraphics[angle=-90]{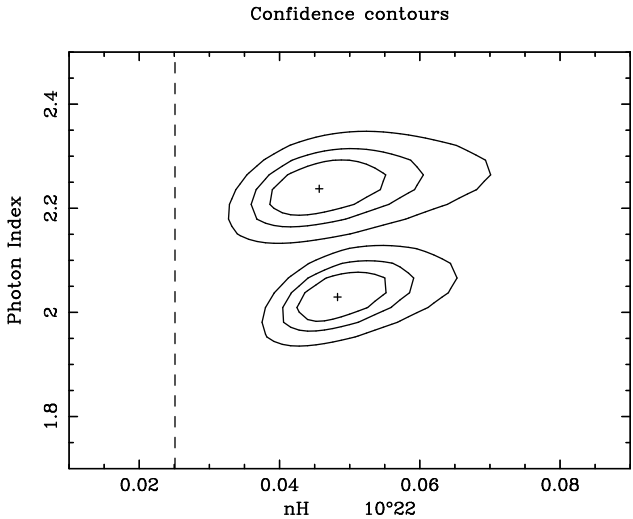}}
\vspace{0.4cm}
\caption{Confidence contours for the photon index and column density
from the single power--law fits to PKS 0548--322 data.
The upper one is for the 7/4/99 observation, 
the lower for the 20/2/99 one. The line indicates the galactic column density.}
\end{figure} 
The planned observation for this object has been split in three parts.
The second observation (26--27/2/99) was very short ($\sim$ 2000 s), 
and only MECS data are available.  For the first observation (20/2/99),
an extraction region of  $6^{\prime}$ of radius has been used for the LECS,
due to the presence of a contaminating source in the field of view, at 
$\sim10^{\prime}$.
We have identified this source with the star \verb+GSC_07061_01558+ 
in the Guide Star Catalog, probably in a flaring 
state: it is much less visible in the MECS image (indicating a very 
soft spectrum) and was completely absent in the LECS and MECS 
images of the third observation.

In both observations with LECS data, a single power--law fit with free 
absorption gives $N_{\rm H}$ values higher than the galactic one, 
at more than 99.9\% confidence (see Fig. 7). 
The spectral index has varied significantly between the three observations, 
not allowing the sum of the datasets to enhance the S/N.
For the third observation the single power--law model fits the data very 
well; instead, the first observation is significantly better fitted by a 
broken power--law (F--test $>99.8\%$), with a break around 4 keV.
In both cases, the data require a column density in excess of the 
galactic one, with high significance (F--tests $>99.9\%$ and $>99\%$ 
respectively, see Table 2 and 3). 
A column density higher than galactic seems also suggested by
the fit with the ``curved" model: the fit with
a galactic $N_{\rm H}$ provides low energies
spectral indices slightly flatter than expected from a peak around 3--4 keV
($\alpha_1=0.4$ at 0.5 keV). 
More likely values are obtained with $N_{\rm H}$
around 3.6--4.5$\times$10$^{20}$ cm$^{-2}$  ($\alpha_1=0.7-0.8$).

Interestingly, this extra absorption could be associated with 
the BL Lac environment.
In fact,
PKS 0548--322 belongs to a rich cluster of galaxies, and has two companions,
one of which is seen in close interaction with the host galaxy of the 
BL Lac object (from NTT images, see Falomo et al. 1995).
Previous X--ray observations with ROSAT and ASCA (see Sambruna et al. 1998)
show some evidence of spectral absorption features around 0.5 keV, 
which suggest the presence of circumnuclear ionized gas, 
perhaps fueled by the tidal interaction.
Unfortunately our data, due to the low S/N ratio in the low energy band, 
cannot confirm nor exclude these features.

PKS 0548--322 was detected  up to $\sim30$ keV by the PDS: 
the data are compatible with the extrapolation of the power--law
from lower energies.

\begin{figure}
\centering
\resizebox{8.2cm}{!}{\includegraphics[angle=-90]{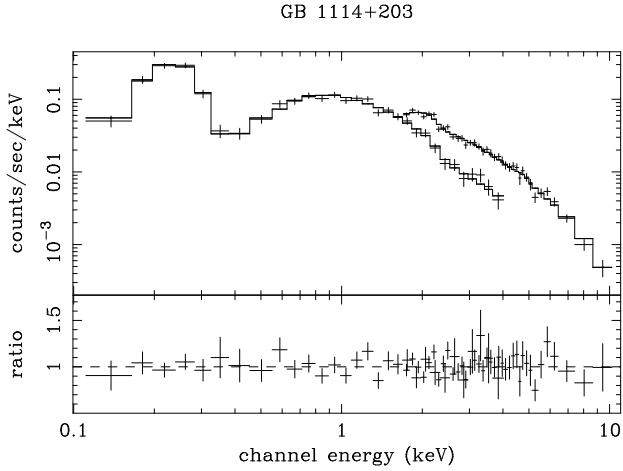}}
\caption{Best fit data and model for GB 1114+203: broken power-law with 
fixed galactic absorption. The source was not detected in the PDS.}
\end{figure}

\subsubsection{GB 1114+203}

This source presents a very steep spectrum, with a further steepening beyond 
1.6 keV best fitted by a broken power--law model.
The column density is in agreement with the galactic value, within the errors. 
No detection was achieved in the PDS instrument. Although the flux level
at 1 keV is comparable with that of other sources in the sample, the steepness
of the spectrum ($\alpha\sim 1.9$ beyond 1.6 keV) makes the source too faint 
to be detected in the PDS range.
The two steep indices locate the synchrotron peak at lower energies 
than the observed X--ray band, thus identifying this object as a typical 
``normal" HBL.

\subsubsection{1ES 1218+304}
As the previous object, also this source presented a steep spectrum
best fitted by a broken power--law with high significance ($>99.9\%$). 
The found column density is in agreement with the galactic value.  
The source is detected in the PDS up to $\sim30$ keV, and the data,
although slightly higher, are compatible with the extrapolation of the
LECS+MECS best fit model within the (large) uncertainties (Fig. 9).
Again, the two steep X--ray indices 
identify this object as a normal HBL.
Previous EXOSAT and ROSAT observations reported similar spectral properties,
with steep slopes near the  \sax values 
($\alpha\approx1.5\pm0.1$ and $\alpha=1.22\pm0.03$ in the ME and PSPC 
instruments, 
respectively; see Sambruna et al. 1994, and Fossati et al. 1998).

\subsubsection{1ES 1426+428}
The spectrum of this object is very well fitted by a single flat power--law 
model, with galactic absorption: there is no clear sign of bending
in the shape of the residuals (see Fig. 10). 
Broken power--law models do not statistically improve the fit, and moreover 
give values of \als and \alh both flatter than unity
[they are also only marginally different (\alh--\als $\sim$0.1)].
The data in the PDS band are particularly important for this
source, in order to assess or not the presence of a steeper component 
($\alpha>1$) that would reveal the position of the synchrotron peak.

\begin{figure}
\centering
\resizebox{8.2cm}{!}{\includegraphics[angle=-90]{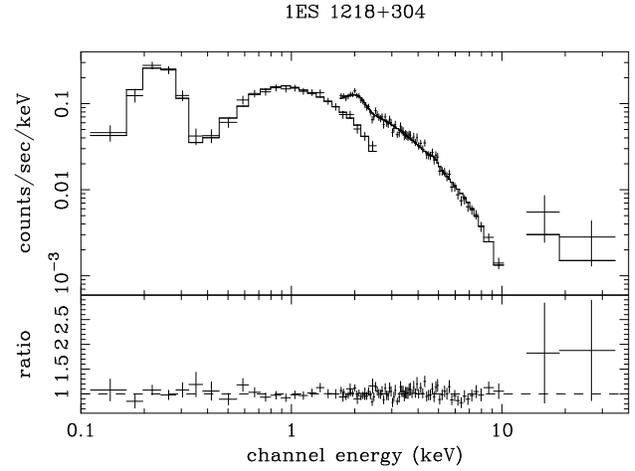}}
\caption{Best fit data and model for 1ES 1218+304: broken power-law with absorption
fixed at the galactic  values.}
\end{figure}

Unfortunately, the PDS data are contaminated by another known
hard source, namely the quasar GB 1428+422 (Fabian et al. 1998).
This object, characterized by a very flat spectral index ($\alpha<0.5$),
lies at $41\arcmin$ from 1ES 1426+428, thus inside the PDS f.o.v. but outside
the MECS f.o.v.. 
For a lucky coincidence however, the quasar was observed by \sax  on 
4/2/99, four days before our observation.
Thanks to the collaboration between the two proposing groups, we have been
able to estimate the relative inter--contamination.
Furthermore, we have also checked in NED and WGACAT databases for other 
potentially contaminating sources within $80\arcmin$ from 1ES~1426+428,
selecting hard sources (HR $>0.8$) and/or with a high countrate,
excluding those already identified as stars,
and estimating their flux in the PDS range under 
the assumption of a power--law spectrum.

\begin{figure*}
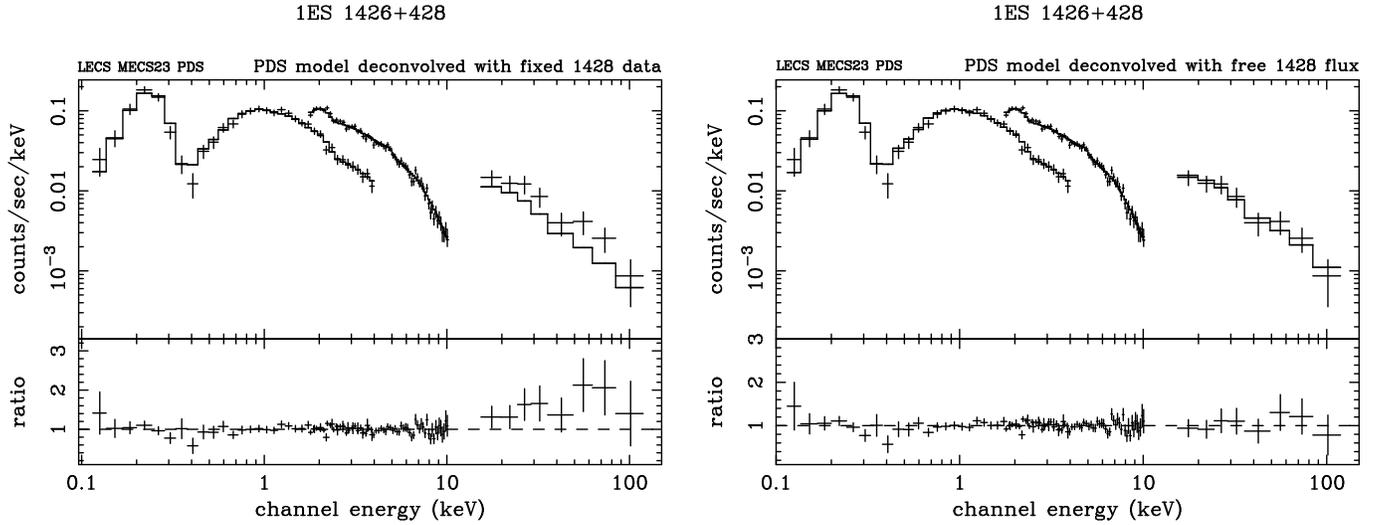

%
%
\resizebox{\hsize}{!}{\includegraphics[angle=-90]
{H2568L10.PS}\hspace{1.5cm}\includegraphics[angle=-90]{H2568R10.PS}}
\caption{Single Power-law fits to the 1ES 1426+428 data. 
The PDS data have been deconvolved  with the
GB 1428+422 normalization either  fixed at \sax value (left panel) or free to vary 
(right panel). In both cases the slope of the GB 1428+422 spectrum is fixed at the
\sax value ($\alpha=0.42$).}
\end{figure*}

Besides the quasar,
the highest contributions came from two sources:
WGA J1426.1+4247 ($\alpha\sim 0.57$, \rosat countrate =0.011 cts/s) 
and CRSS 1429.7+4240 ($\alpha\sim 0.87$, \rosat countrate =0.015 cts/s).
Their estimated high energy X--ray flux turned out to be
roughly two orders of magnitude lower than 1ES 1426+428 in the PDS band,
so we did not consider them, leaving GB 1428+422 as the only
contaminating source.

The PDS data of both blazars were contaminated one by the other.
The strategy to disentangle the two contributions has been
the following.
First we fitted only the LECS+MECS data of both sources.
Then we extrapolated these best fit models into the PDS range, for estimating
the relative contaminations. 
We fitted the combined LECS+MECS+PDS data of one source, 
adding in the PDS range
the LECS+MECS model of the other source, after accounting for the instrument 
off--axis response, and cross--checked the results for consistency.
In Fig. 11 the single components of the models used in the 
1ES 1426+428 fit are shown, together with the total expected flux
in the PDS instrument. 
Of course, this technique is based on the assumption that no 
large changes of flux or spectral index occurred in the four days separating 
the two observations, but in any case it gives a reference point 
to estimate possible variations. 
Fig. 10 (left panel) shows the fit including for the PDS model 
the \sax~ spectrum  of GB 1428+422 
($\alpha= 0.42$, F$_{\rm 1~keV}= 0.30\mu$Jy).
The PDS points remain slightly above the flat ($\alpha_x=0.92$) power--law
that fits the 1ES 1426+428 LECS$+$MECS data, 
however the fit is good ($\chi^2_r= 1.06$), with no sign 
of declining slope up to 100 keV. 
A broken power--law model gives even flatter indices above 10 keV
($\alpha_2\approx0.3-0.4$),
however, these values are not very plausible, since,
even not considering the GB 1428+426 contribution, the emerging of such a 
flat component would have implied a 3$\sigma$ detection
in the PDS also in the 100--200 keV band, not observed
($\sim8.2\times10^{-2}$ cts/s against the observed $2.2\pm2.7$ 
$\times10^{-2}$ cts/s).

\begin{figure}
\resizebox{7.5cm}{!}{\includegraphics[angle=-90]{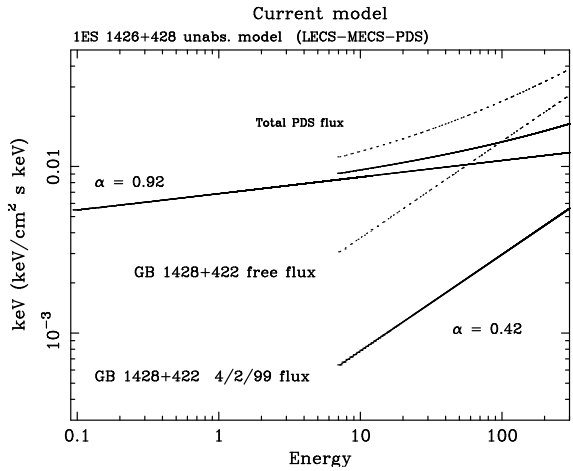}}
\caption{The two models used to fit 1ES 1426+428 PDS data (shown in Fig. 10).
Together with the model for the pointed 
source (the power--law with $\alpha=0.92$), the off-axis GB 1428+422 
contribution 
is shown, with the two different flux levels used. The lower one (thick line) refers to 
the \sax observation value, the upper (dashed line) to the level resulting from the best 
fit obtained with the quasar flux as free parameter. 
The two curved lines correspond to the total flux in the PDS range,
i.e. the sum of the $\alpha=0.92$ power--law with each of the two GB 1428+422
flux levels (thick and dashed lines respectively). } 
\end{figure}

Assuming also the possibility of a large flux variation for GB 1428+422
(not unlikely: in 1998, during pointed ROSAT HRI observations,
it has doubled its flux over a timescale of two weeks or less, 
see Fabian et al. 1999), we also fitted the PDS data letting the 
GB 1428+422 normalization free to vary (Fig. 10, right panel).  
Again, the 1ES 1426+428 spectrum remains flatter than unity up to 100 keV, even 
accounting for a GB 1428+422 increase of a factor $\sim$4 in 4 days, from
$F_{\rm 1~keV}=0.30\;\mu$Jy to $F_{\rm 1~keV}=1.44\;\mu$Jy.
Given the SED of 1ES 1426+428 (see Fig. 14), this result 
quite reliably {\it constrains the synchrotron peak to lie near 
or above 100 keV}.
\begin{figure}
\resizebox{8.2cm}{!}{\includegraphics[angle=-90]{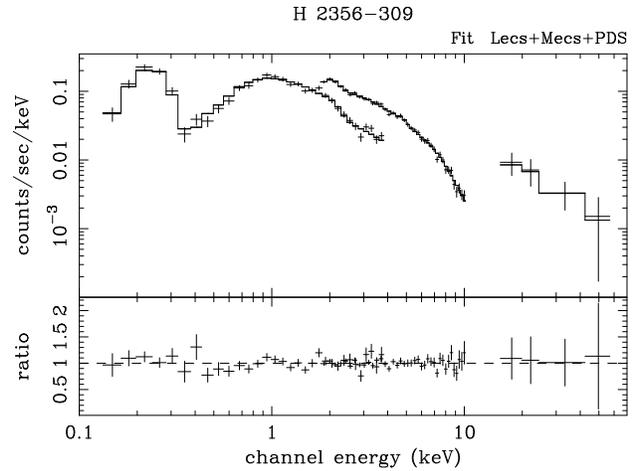}}
\caption{Broken power--law fits to the LECS+MECS+PDS data of H 2356--309.
Absorption fixed at the galactic value.}
\end{figure}

\subsubsection{H 2356--309}

The spectrum of this source is not compatible with a single
power--law model: there is clear evidence for a convex spectral shape,
that is best fitted by a broken power--law with high significance
($>99.9\%$, see Table 3). This locates the synchrotron peak around 1.8 keV.
The found column density is around the galactic values.
This source has been detected also in the PDS instrument up to $\sim50$
keV (see Fig. 12) and the data lie on the extrapolation
of the MECS power--law.

\section{Comparision with previous X--ray observations}

Among the seven sources of our sample, PKS 0548--322 and 1ES 1426+428
have been observed several times also by previuos X--ray missions.
Here we compare our \sax results with older X--ray spectra, in order 
to reveal and study possible variations of the synchrotron peak frequency. 

\subsection{PKS 0548--322}
Variations of the location of the synchrotron peak have been observed
even during our \sax observations,
showing a convex spectrum peaked around $\sim 4$ keV in the first 
dataset and a steep spectrum in the third observation (locating the 
synchrotron peak below the observed band).  
The ASCA observations (Sambruna et al. 1998)
confirm the presence of a convex spectrum, but with the break at slightly 
lower energies, around 1.7 keV ($\alpha_{\rm soft}=0.9\pm0.1$ and
$\alpha_{\rm hard}=1.04\pm0.02$). 
The ROSAT slope agrees with the lower energy index of the ASCA 
broken power--law ($\alpha=0.84\pm0.05$, Sambruna et al. 1998).
The EXOSAT and GINGA satellites observed the source
three and five times respectively, showing again an 
interesting spectral variability.
The EXOSAT observations of 2 November 1983 (Garilli \& Maccagni 1990) and
7 March 1986 (Barr et al. 1988) report a ``convex" spectrum 
similar to our first dataset, with break energies indicating a 
synchrotron peak at $2.8\pm0.8$ and $\sim 5$ keV, respectively.
The 30 Nov. 1983 observation, instead, is well fitted by a steep single 
power--law ($\alpha=1.2^{+0.1}_{-0.2}$ in the 0.02--7.2 keV band), as our 
third dataset (Garilli \& Maccagni 1990).
Four out of the five GINGA observations (Tashiro et al. 1995) 
carried out between January 1990 and March 1991,
showed an X--ray slope $\alpha_x\sim 1$ between 2--30 keV
(hence the synchrotron peak lies in this range).
One observation instead (19 Feb. 1991), quite interestingly, 
was characterized by a very flat spectral index ($\alpha=0.84\pm0.02$) 
over the whole energy range, which would locate the synchrotron peak at 
energies $\geq$30 keV.
These data suggest a possible behavior similar to Mkn 501, with the
synchrotron peak which sometimes reaches the hard X--ray band, 
even if most of the time is rather stable in the range 1--5 keV. 

PKS 0548--322 has been observed also in the TeV band, but 
so far no detection was reported: CANGAROO observations  
between October and November 1997 showed no evidence for an 
on--source excess (Roberts et al. 1999). 
The simultaneous optical and XTE All Sky 
Monitor X--ray measurements did not show any special activity.

\subsection{1ES 1426+428}

\begin{figure}
\centering
\resizebox{8cm}{!}{\includegraphics[angle=-90]{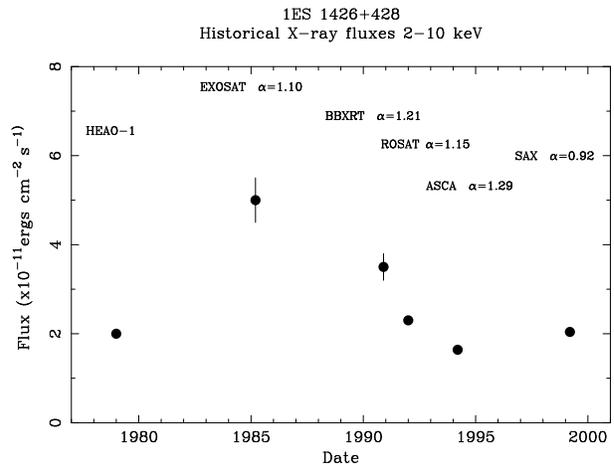}}
\caption{Historical X--ray fluxes (2--10 keV) measured by the satellites 
 HEAO--1, EXOSAT,  ROSAT,  ASCA, and the BBXRT experiment (see Sambruna et al. 1997).
The fluxes have been obtained from  single power--law fits with galactic 
absorption to
the 2--10 keV continuum (except ROSAT, where the values are from the 
extrapolation in the 2--10 keV band of the best fit model).
In the upper part are also displayed the corresponding spectral indices. 
The errors are estimated from those on normalization and spectral index, at 90\%
confidence level. For ROSAT, ASCA and \sax, they are comparable with the marker size.
Note how in the \sax observation 
the  flux level was similar to the ASCA and ROSAT ones, but with a very
different spectral index, indicating a shift in the peak frequency 
of about two orders of magnitude.}
\end{figure}

Besides ROSAT, this source has been previously observed also by 
HEAO--1, EXOSAT, ASCA and the BBXRT experiment 
(details in Sambruna et al. 1994, 1997). 
In Fig. 13 we report the historical flux levels in the 2--10 keV band, 
together with the measured spectral indices in that band.
In all the previous observations, there is no evidence of a flat spectrum
above 2 keV, even in the presence of large flux variations.
The \sax observation is the first where this source showed a flat spectrum 
up to 100 keV.
In BBXRT and ASCA data, the full band spectrum is 
best fitted by a broken power--law: in this case, the resulting 
break energy locates the synchrotron peak, for both observations, between 
1 and 2 keV.
This source therefore has undergone a shift in the peak frequency of at 
least two orders of magnitude.  
Note that, as shown by Fig. 13, the \sax flux in the
2--10 keV band of 1ES 1426+428 is not particularly high: 
it is the second lowest state observed, about $\sim$20\% 
higher than the lowest one, observed by ASCA ($2.04$ vs. $1.64$ \ergs{-11}, 
in the 2--10 keV band).
This indicates that the object 
can be very powerful above 10 keV while remaining inconspicuous
in the soft X--ray band.
This is also supported by the XTE monitoring,
which did not show any special activity during the \sax observation.
The other two ``over 100 keV" sources (i.e. Mkn 501
and 1ES 2344+514) showed a markedly different behavior, becoming
very active and strong also below 10 keV when their synchrotron 
spectrum peaks at high energies.
1ES 1426+428 has shown that the soft X--ray activity is not 
always necessary to be a strong 100 keV synchrotron emitter.
At the same time, in this source, a relatively faint 2--10 keV state does
not guarantee a hard spectrum (see e.g. the ASCA and ROSAT
data in Fig. 13): all this is revealing of a complex 
relation between the injected power and the location of the 
synchrotron peak.
\begin{figure*}
\centering
\resizebox{15.4cm}{!}{\includegraphics[angle=0]{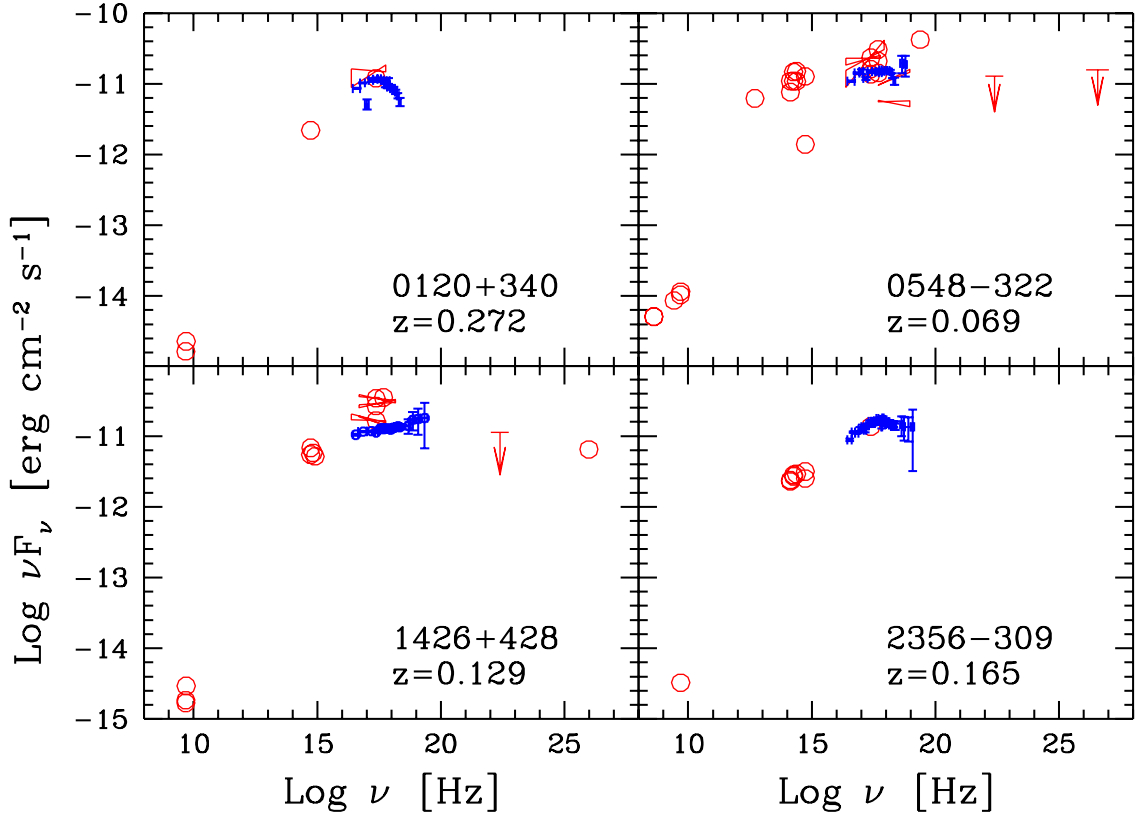}}
\vspace{-4 cm}
\caption{The SEDs of the 4 BL Lacs observed in AO2, made with {\it Beppo}SAX 
(small filled points)
and literature data (open circles, bow-ties for  
slope informations and  arrows for upper limits).  
All four sources peak in the X--ray band. }
\end{figure*}

\section{Discussion}
\begin{figure*}
\centering
\vspace*{2mm}
\resizebox{16cm}{!}{\includegraphics[angle=-90]{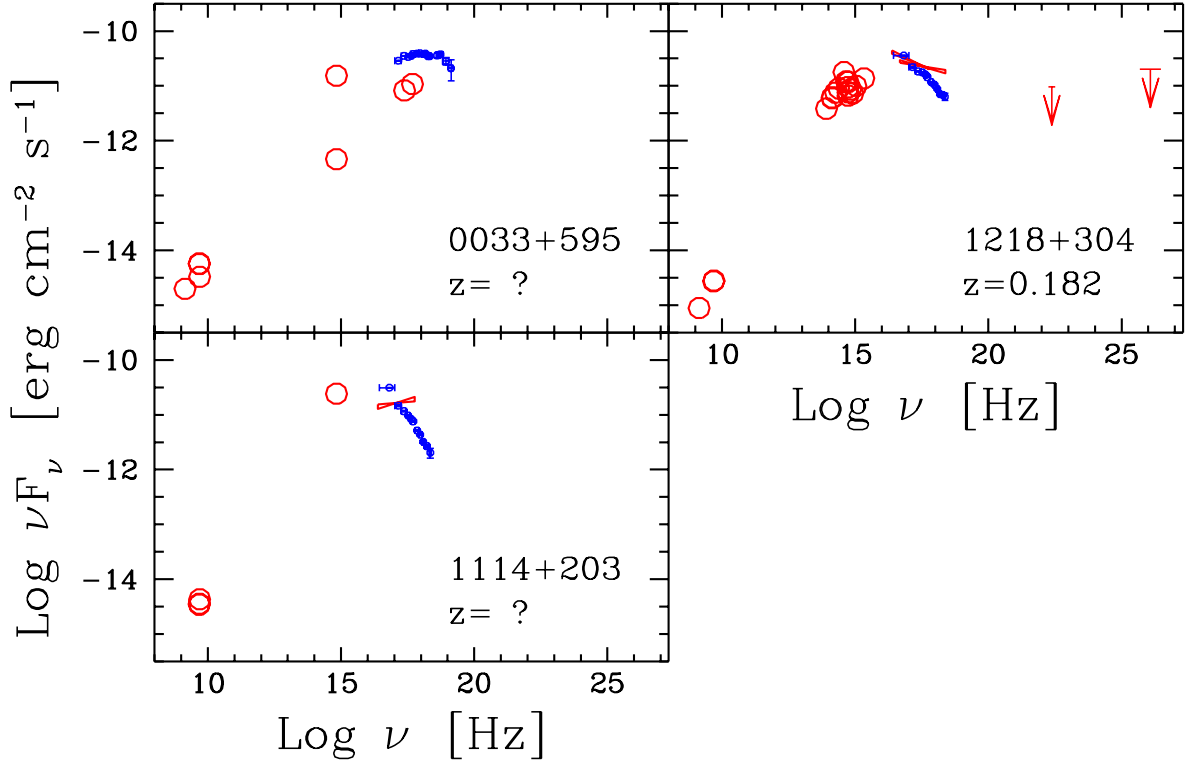}}
\vspace*{-1.4cm}
\caption{The SEDs of the 3 BL Lacs observed in AO3, made with 
{\it Beppo}SAX and literature data.
Only 1ES 0033+595 peaks in the observed X--ray band; the other two objects 
peak at  lower energies. A polynomial fit as in Wolter et al. 1998 
( log($\nu F_{\nu}$)=$a+b$(log$\nu$)$+c$(log$\nu$)$^2+d$(log$\nu$)$^3$) gives 
log$(\nu_{peak})$=15.4 and 16.3 for GB1114+203 and 1ES 1218+304, respectively. }
\end{figure*}

The main purpose of our observations was to characterize the global
spectral properties of the selected candidates, verifying their 
``extremeness" by 
locating the position of the synchrotron peak by means of the X--ray spectrum
itself and by the shape of the overall SED.
The SEDs of the 7 sources, constructed with our \sax and 
literature data, are shown in Fig. 14 and 15. 
The \sax observations have revealed and confirmed 
the ``extreme" nature of five  objects: 
for four of them (1ES 0033+595, 1ES 0120+340, PKS 0548--322 and H 2356--309) 
the X--ray data themselves show the peak of the synchrotron emission,
presenting a convex spectrum best fitted with a broken power--law 
whose break locates the peak in the 1--5 keV range.
For 1ES 1426+428, instead, the flat X--ray spectrum up to 100 keV
constrains the synchrotron peak to lie near or even above that value,
and makes this source the third object ever found with such high peak 
frequencies, after Mkn 501 and 1ES 2344+514.
For the last two sources (GB 1114+203 and 1ES 1218+304), again 
best fitted by a broken power--law, the indices are both steep, thus locating 
the peak below the observed X--ray band (see Fig. 15).
Their steep spectrum qualifies them as typical HBLs,
for which the  X--ray band is dominated by the tail of the synchrotron 
emission after the peak.
These results confirm that the criteria adopted to select
our candidates were highly efficient (5 out of 7), and we are beginning to 
populate the high energy end of the synchrotron peak sequence.

An important aspect that our observations have pointed out, compared with 
older data, is the
high variability of the peak frequency exhibited by these objects:
the observed range of $\nu_{\rm peak}$ variations has been from less than 
$0.1$ to 1.4 keV for 1ES 0120+340, from 
$<0.1$ to $>20$ keV for PKS 0548--322, and from 
$\sim$1 to over 100 keV for 1ES 1426+428. 
This can affect the 
classification of an object as ``extreme": in fact,
if observed only once by \sax, 1ES 0120+340 and PKS 0548--322 would have been 
classified as normal HBLs (see Table 2). 
Mkn 501 and 1ES 2344+514 too exhibited HBL properties before their 
observations during flaring states. 
It is likely, therefore, that other ``extreme" objects could be hidden
in the HBL class, and would be uncovered if observed many times.
Repeated observations are therefore necessary to properly identify 
this type of objects, and also to sample their spectral behavior in 
connection with flux changes. 
Flux variations, however, should not be the only criteria to trigger
the observations for the most extreme states: 1ES 1426+428 proves that 
the shift in the peak frequency can be de--coupled from high flux states
in the soft X--ray band.

A modeling of the SEDs and the resulting physical parameters
will be presented in a forthcoming paper (for a preliminary Synchrotron 
Self Compton model to the AO2 sources, see Fig. 1 of Costamante et. al. 1999). 
Although more objects are necessary to draw reliable and statistically
significant conclusions, it is nevertheless interesting to start 
analyzing the characteristics of these extreme objects also in the context 
of the entire class of BL Lac objects.

\begin{figure}
\centering
\resizebox{8cm}{!}{\includegraphics[angle=-90]{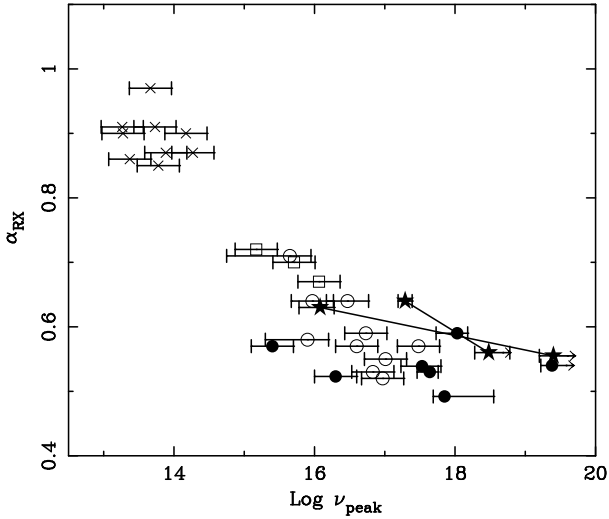}}
\caption{The radio to X--ray broad band spectral index 
$\alpha_{rx}$ as a function of $\nu_{\rm peak}$.
Open symbols correspond to data from Comastri et al. 1995 (crosses and squares),
and Wolter et al. 1998 (circles). 
Filled circles are the sources analyzed in this paper. 
The stars represent the quiescent and flaring states (connected by a line)
for 1ES 2344+514 and Mkn 501. 
The $\alpha_{rx}$ parameters have been calculated as in Wolter et al. (1998),
as well as the $\nu_{peak}$, when not directly measured in the X--ray spectrum.
 }
\vspace*{-3mm}
\end{figure}

Fig. 16 shows the broad band spectral index $\alpha_{rx}$ as a function 
of the frequency of the synchrotron peak,
for the  BL Lacs analyzed in this paper plus Mkn 501 and 1ES 2344+514, 
together with the HBLs analyzed in Wolter et al. (1998) and
the RBLs analyzed in Comastri et al. 1995. 
It can be seen that the almost linear correlation between $\alpha_{rx}$ 
and $\log \nu_{\rm peak}$ flattens out for large values of $\nu_{\rm peak}$,
reaching a ``stable" value of $\alpha_{rx}\sim 0.5$.
This is in agreement with the scenario of a synchrotron peak moving smoothly 
from lower to higher energies: when the peak moves beyond the 
X--ray band, both the radio and the X--ray fluxes are produced by 
the same branch of the synchrotron emission, 
and so both fluxes change similarly as the peak moves at still 
higher frequencies. 
Consequently $\alpha_{rx}$ stabilizes at a (flat) value.
Instead, when the radio and X--ray fluxes are produced by different 
branches of the synchrotron emission (before and after the peak), 
the radio and the X--ray fluxes change differently as the peak shifts, 
thus changing $\alpha_{rx}$.

The ``moving peak scenario" also nicely accounts for the 
relation between $\alpha_{x}$ and $\nu_{\rm peak}$ 
(Padovani \& Giommi 1996), shown in Fig. 17.
To reliably compare our results with previous ROSAT results,
in Fig. 17 we plot $Beppo$SAX spectral indices in the soft band
(i.e. the first spectral index for a broken power-law model).
We can see that LBLs are characterized by flat indices ($\alpha_x<1$),
since their X---ray emission is likely dominated by the flat inverse
Compton process.
Intermediate BL Lacs, with $\nu_{\rm peak}\sim 10^{14}$--$10^{15}$ Hz,
are characterized by $\alpha_{x}\sim 1$.
For them, broad band X--ray observations should reveal a concave spectrum,
with the soft X--rays dominated by the steep tail of the synchrotron
emission, and with the flat inverse Compton radiation emerging at higher
energies (see the nice example of ON 231 in Tagliaferri et al. 2000).
The steepest indices correspond to HBLs, where the X--rays are 
entirely dominated by the steep tail of the synchrotron emission.
But as $\nu_{\rm peak}$ increases further, the synchrotron peak
moves into the X--ray band, making the spectrum to flatten again,
eventually reaching a ``stable" value ($\alpha_x\sim 0.5$)
corresponding to the typical slope of the synchrotron spectrum 
well before its peak. 
These are the objects that we have called extreme BL Lacs in this paper
(i.e. with $\nu_{\rm peak}>1$keV).

\begin{figure}
\centering
\resizebox{8cm}{!}{\includegraphics[angle=-90]{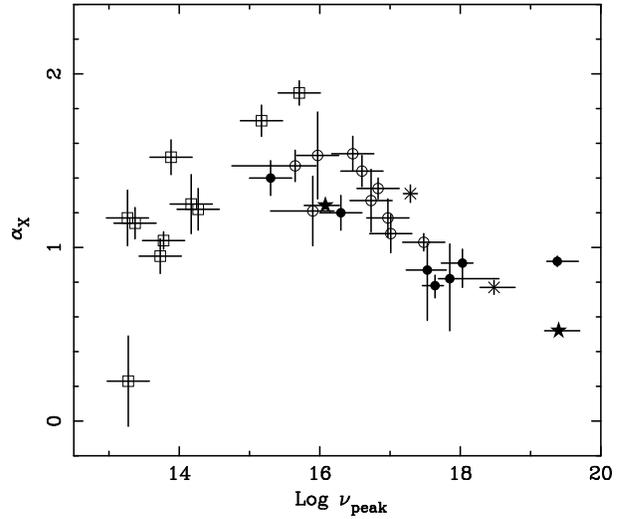}}
\caption{The X--ray spectral index $\alpha_{x}$ as a function of 
$\nu_{\rm peak}$ (Padovani \& Giommi 1996).
Open squares and circles are from Comastri et al. (1995)  
and Wolter et al. (1998).
The values from Comastri et al. (1995) are obtained from ROSAT observations
(for \sax results on LBLs, see Padovani et al. 2001).
Filled circles mark the BL Lacs analyzed in this paper. 
For consistency, the $\alpha_{x}$ here reported
is the soft one, in case of broken power--law best fits.
The stars (Mkn 501) and asterisks (1ES 2344+514) refer to the quiescent and 
flaring states values for these two sources.}
\end{figure}

The left panel of Fig. 18 shows $\nu_{\rm peak}$ as a function of the 
radio luminosity ($\nu L_{\nu}$ at 5 GHz) for our targets and
for a few samples of blazar objects (1Jy BL Lacs, Slew Survey BL Lacs and 2Jy
FSRQs), including the objects studied by 
Wolter et al. (1999) as well as Mkn 501 and 1ES 2344+514 in the flaring state.
The solid line is the phenomenological relation proposed by Fossati et al. 
(1998) for the analytic representation of the general SED.
In the context of the samples under considerations, 
our extreme BL Lacs, having luminosities similar to HBLs but larger 
$\nu_{\rm peak}$, indicate 
a wider range of peak frequencies covered by HBLs, compared to lower peaked
objects. 
Indeed, the synchrotron peak frequency of the extreme BL Lacs and some HBLs 
is highly variable (as in the case of Mkn 501, 1ES 2344+514 and
1ES 1426+428, see Fig. 18). 

At present, we do not know if this extreme spectral variability
is a characteristic common to all blazars or if it is a peculiarity of
only the HBLs (or some of them).
It is however interesting to note that for some well studied LBL and
FSRQ the value of $\nu_{\rm peak}$ seems to remain  more constant. 
As an illustration we show in Fig. 18 (right panel) the observed range of variation of both 
$\nu_{\rm peak}$ and $L_{\rm peak} \equiv \nu_{\rm peak} L_{\nu_{\rm peak}}$
for three ``low peak" sources of different luminosity 
(BL Lac itself, OJ 287 and 3C 279).
The high $\nu_{\rm peak}$ objects seem to be characterized by a much larger 
$\nu_{\rm peak}$ variability, compared to the others, given
more or less the same range of luminosity variations.

The high synchrotron peak frequencies which characterize, by definition, the 
extreme BL Lac class, flagging the presence of high energy electrons, make these
sources good candidates for TeV emission through the inverse Compton process,
as also indicated by the TeV detections of Mkn 421, Mkn 501 and 1ES 2344+514 
(Catanese et al. 1997, 1998; see also Stecker, de Jager \& Salamon 1996).
\begin{figure*}
\centering
\vspace*{-2cm}
\hspace*{-0.5cm}
\resizebox{19cm}{!}{\includegraphics[angle=-90]{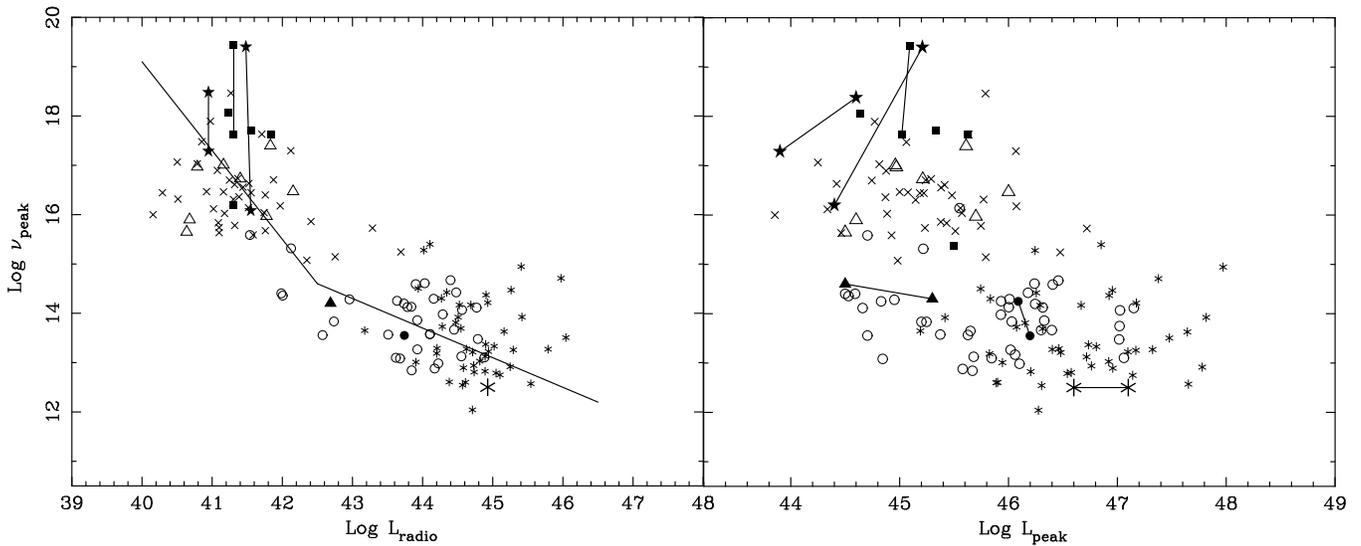}}
\vspace*{-3cm}
\caption{Peak frequency vs. Radio luminosity at 5 GHz,
and  vs.  luminosity at the peak frequency
for the extreme BL Lacs data (filled squares) and the HBL data from Wolter et
al. 1998 (open triangles). The other data are from Fossati et al. 1998
(1 Jy and Slew Survey BL Lacs,  and 2Jy FSRQs samples, marked as circles, 
crosses and asterisk respectively).
The short lines connect different observations parameters for the same source:
on the left,  Mkn 501, 1ES 2344+514 and 1ES 1426+428. 
On the right, together with the previous three sources, two states for 
three objects of different luminosity are also reported, for comparision: 
BL Lac itself (an Intermediate--LBL object, filled triangles),
OJ287 (a LBL, filled circles) and 3C279 (a typical FSRQ, large asterisk).
On the left,
the broken line shows the relation between $\nu_{\rm peak}$ and L$_{\rm radio}$
assumed by Fossati et al. 1998 ($\nu_{\rm peak}\propto $L$^{-1.8}$ and
$\nu_{\rm peak}\propto $L$^{-0.6}$ before and after logL$_{\rm radio}$=42.5,
respectively).  
For all the objects except the  ``extreme",
the frequency of the peak is evaluated from polynomial fits of the SEDs,
and are uncertain of a factor 2 or 3.
Therefore the variations here reported should be considered as indicative
of the typical ``box" of parameters covered by the single source.
Note how the high peak objects seem to show a higher variability
of synchrotron $\nu_{\rm peak}$, while lower peaked
sources (i.e. 3C279) show a rather steady $\nu_{\rm peak}$ among different
luminosity states.  }
\end{figure*}
A significant role in the propagation of such high energy photons, however, is 
played by the absorption resulting from pair production interactions
with the intergalactic infrared background. In fact, it has been shown that
photons with energies greater than $\sim1$ TeV should be removed by the spectra
of sources with redshifts $>0.1$, according to the estimates of the
intergalactic IR background (Stecker \& De Jager 1997, 1998).
In this respect, 
the extreme BL Lacs could be very useful as probes for the IR
background, thanks to their different redshifts which make possible a good 
sampling of the optical dephts along $z$.
1ES 1426+428, in particular, should be very interesting, given its
high synchrotron frequencies (in the ``Mkn 501 class") and its redshift 
at the border of the expected observable region ($z=0.129$).

\begin{acknowledgements}
This research has made use of the NASA/IPAC Extragalactic Database (NED) 
which is operated by the Jet Propulsion
Laboratory, California Institute of Technology, under contract with 
the National Aeronautics and Space
Administration. We thank the {\it Beppo}SAX Science Data Center for 
their support in the data analysis. 
L.C. thanks the Cariplo Foundation and the Italian Space Agency for support.
 
\end{acknowledgements}


\begin{thebibliography}{}
\bibitem[]{} Balucinska-Church M. \& McCammon D., 1992, ApJ, 400, 699
\bibitem[]{} Barr P., Giommi P., Maccagni D., 1988, ApJ, 324, L11
\bibitem[]{} Boella G., Butler R.C., Perola G.C., Piro L., Scarsi L., 
Bleeker J.A.M., 1997, A\&AS, 122, 299
\bibitem[]{} Brinkmann W. \& Siebert J., 1994, A\&A, 285, 812
\bibitem[]{} Caccianiga A., Maccacaro T., Wolter A.,
 Della Ceca R., Gioia I.M., 1999, ApJ, 513, 51
\bibitem[]{} Catanese M., Bradbury S.M., Breslin A.C. et al. 1997, ApJ, 487L, 143
\bibitem[]{} Catanese M., Akerlof C.W., Badran H.M. et al. 1998, ApJ, 501, 616
\bibitem[]{} Comastri A., Molendi S. \& Ghisellini G. 1995, MNRAS, 277, 297
\bibitem[]{} Costamante L., Ghisellini G., Giommi P. et al. 1999, New Extreme 
Synchrotron BL Lac Objects. In the proceedings of the conference ``X--ray Astronomy '99",
Bologna, Italy, September 1999. astro-ph/0001410
\bibitem[]{} Dermer C.D., Sturner S.J. \& Schlickeiser R., 1997, ApJS, 109, 103
\bibitem[]{} Dickey J.M. \& Lockman F.J., 1990, ARA\&A, 28, 215
\bibitem[]{} Dondi L. \& Ghisellini G., 1995, MNRAS, 273, 583
\bibitem[]{} Elvis M., Lockmann F.J., Wilkes B.J., 1989, AJ, 97, 777
\bibitem[]{} Fabian A.C.,  Iwasawa K., McMahon R.G., Celotti A.,
  Brandt W.N., Hook M., 1998, MNRAS, 295L, 25
\bibitem[]{} Fabian A.C., Celotti A., Pooley G., Iwasawa K., Brandt W.N.,
  McMahon R.G., Hoenig M.D., 1999, MNRAS, 308L, 6
\bibitem[]{} Falomo R. \& Kotilainen J.K.  1999, A\&A, 352, 85
\bibitem[]{} Falomo R., Pesce J. E., Treves A., 1995, ApJ, 438L, 9
\bibitem[]{} Fiore F., Guainazzi M., Grandi P. 1999,
             Cookbook for NFI {\it Beppo}SAX Spectral Analysis v. 1.2,
             available at www.sdc.asi.it
\bibitem[]{} Fossati G., Celotti A., Ghisellini G., Maraschi L., 1997, MNRAS, 289, 136
\bibitem[]{} Fossati G., Maraschi L., Celotti A., Comastri A., Ghisellini G., 
1998, MNRAS, 299, 433
\bibitem[]{} Fossati G., Celotti A., Chiaberge M. et al.  2000, ApJ, 541, 166
\bibitem[]{} Garilli R. \& Maccagni D., 1990, A\&A, 229, 88
\bibitem[]{} Gehrels N. 1986, ApJ, 303, 336
\bibitem[]{} Ghisellini G., Maraschi L. \& Dondi L., 1996, ApJS, 120, 153
\bibitem[]{} Ghisellini G. \& Madau P., 1996, MNRAS, 280, 67
\bibitem[]{} Giommi P., Padovani P., Perlman E. 2000, MNRAS, 317, 743
\bibitem[]{} Laurent--Muehleisen S.A., Kollgaard R.I., Ciardullo R., Feigelson E.D., 
Brinkmann W.,  Siebert J., 1998, ApJS, 118, 127
\bibitem[]{} Morrison R. \& McCammon D., 1983, ApJ, 270, 119
\bibitem[]{} Padovani P. \& Giommi P.,1995, ApJ, 444, 567
\bibitem[]{} Padovani P. \& Giommi P.,1996, MNRAS, 279, 526
\bibitem[]{} Padovani et al., 1998, proceedings of the ESO/ATNF Workshop ``Looking 
Deep in the Southern Sky", Sydney, Australia, Dec. 1997, astroph/9803298
\bibitem[]{} Padovani P., Costamante L., Giommi P. et al., 2001, A\&A, submitted
\bibitem[]{} Perlman E., Padovani P., Giommi P., Sambruna R., Jones L.R., Tzioumis A.,
 Reynolds J., 1998,  AJ, 115, 1253
\bibitem[]{} Pian E., Vacanti G., Tagliaferri G. et al. 1997, ApJL, 492, 17
\bibitem[]{} Roberts M.D., McGee P., Dazeley S.A. et al., 1999, A\&A, 343, 691
\bibitem[]{} Sambruna R.M., Barr P., Giommi P., Maraschi L.,
Tagliaferri G., Treves A., 1994, ApJS, 95, 371
\bibitem[]{} Sambruna R.M., Maraschi L. \& Urry C. M., 1996, ApJ, 463, 444
\bibitem[]{} Sambruna R.M., George I.M.,
Madejski G., Urry C.M., Turner T.J., Weaver K.A., Maraschi L., Treves A., 
1997, ApJ, 483, 774
\bibitem[]{} Sambruna R. M. \% Mushotzky R. F., 1998, ApJ, 502, 630
\bibitem[]{} Scarpa R., Urry C.M., Falomo R., Pesce J., Webster R., O'Dowd M., Treves A.,
1999, ApJ, 521, 134
\bibitem[]{} Sikora M., Begelman M.C. \& Rees M., 1994, ApJ, 421, 153
\bibitem[]{} Stecker F.W., de Jager O.C. \&  Salamon M. H., 1996, ApJ, 473L, 75 
\bibitem[]{} Stecker F.W. \& de Jager O.C., 1997, ApJ, 476, 712
\bibitem[]{} Stecker F.W. \& de Jager O.C., 1998, A\&A, 334L, 85
\bibitem[]{} Stocke J.T., Morris S.L., Gioia I.M., Maccacaro T., Schild R., 
Wolter A., Fleming T.A., Henry J. P.  1991, ApJS, 76, 813
\bibitem[]{} Tagliaferri G.,  Ghisellini G., Giommi P. et al., 2000, A\&A, 354, 431 
\bibitem[]{} Tashiro M., Makishima K., Ohashi T., Inda-Koide M., Yamashita A., Mihara T.,
Kohmura Y., 1995, PASJ, 47, 131
\bibitem[]{} Urry C. M. \& Padovani P., 1995, PASP, 107, 803
\bibitem[]{} Wolter A., Comastri A., Ghisellini G. et al. 1998, A\&A, 335, 899
\end{thebibliography}
\end{document}